\documentclass[traditabstract,rnote]{aa}
\usepackage{graphicx}
\usepackage{natbib}
\usepackage{txfonts}
\def\ltapprox{\raise 2pt \hbox {$<$} \kern-1.1em \lower 5pt \hbox {$\approx$}}

\def\ltsim{\; \raise0.3ex\hbox{$<$\kern-0.75em \raise-1.1ex\hbox{$\sim$}}\; }
\def\gtsim{\; \raise0.3ex\hbox{$>$\kern-0.75em \raise-1.1ex\hbox{$\sim$}}\; }

\def\eg{{\it e.g.,~}}

\begin{document}
\title{Constraining relativistic protons and magnetic fields in galaxy
clusters through radio \& $\gamma$-ray observations : the case of A2256}

\author{Gianfranco Brunetti\inst{1}\fnmsep\thanks{\email{brunetti@ira.inaf.it}}}

\authorrunning{G. Brunetti}

\offprints{G. Brunetti}

\institute{INAF - Istituto di Radioastronomia, via P. Gobetti 101,I-40129 
Bologna, Italy}

\date{Received...; accepted...}

\abstract{
Giant radio halos are the most relevant examples of diffuse 
synchrotron emission from galaxy clusters. 
A number of these sources have very steep spectrum, with spectral 
index $\alpha \geq 1.5-1.6$ ($F(\nu) \propto \nu^{-\alpha}$), and are 
ideal targets to test current 
models for the origin of the relativistic particles.
A2256 hosts the nearest radio halo with very steep spectrum, 
with $\alpha =1.61$, and a very large population of relativistic protons 
in the cluster would be necessary to explain the halo as due to 
synchrotron emission from secondary particles.
In this case the 0.1-1 GeV $\gamma$--ray luminosity is expected 10-20 
times larger than that of clusters hosting radio halos with similar 
radio power at GHz frequencies but with spectra more typical of the 
presently observed halo population, $\alpha \sim 1.2$. Under these 
assumptions incoming FERMI/GLAST observations are expected to detect A2256, 
provided that the magnetic field in the central cluster region 
is $\leq$10-15 $\mu$G. We show that this will allow for a prompt test of 
hadronic models for the origin of radio halos, and for complementary 
constraints on both the cluster magnetic field and the physics of particle 
acceleration mechanisms.}


\maketitle

\section{Introduction}

Galaxy clusters are the largest gravitationally bound objects 
in the Universe.
During cluster mergers energy may be channelled into the amplification of the
magnetic fields (Dolag et al.~2005; Ryu et al.~2008)
and into the acceleration of relativistic primary
electrons (CRe) and protons (CRp) via shocks and turbulence
(\eg Ensslin et al.1998; Sarazin 1999;
Blasi 2001; Ryu et al. 2003; Gabici \& Blasi 2003;
Pfrommer et al. 2006; Brunetti \& Lazarian 2007).
CRp have long life-times and remain confined within
clusters for a Hubble time (V\"olk et al. 1996;
Berezinsky et al. 1997; Ensslin et al. 1997), they are expected to 
be the dominant 
non-thermal particle component in the ICM and should produce 
secondary particles due to collisions with thermal 
protons (\eg Blasi et al.~2007 for review).

\noindent
Direct evidence for magnetic fields and relativistic particles, 
mixed with the 
thermal Intra-Cluster-Medium (ICM), comes from  
radio observations that detect Mpc-sized diffuse radio 
sources, radio halos and relics, in a fraction of X-ray luminous
galaxy clusters in merging phase (\eg Ferrari et al.~2008 for review).
Extended and fairly regular diffuse synchrotron emission, in the
form of giant radio halos, may be produced
by secondary electrons injected during proton-proton collisions
(hadronic models, e.g. Dennison 1980; Blasi \& Colafrancesco 1999;
Pfrommer \& Ensslin 2004), or assuming that relativistic electrons 
are re-accelerated in-situ by MHD turbulence generated in the 
ICM during cluster-cluster mergers (re-acceleration models, e.g.
Brunetti et al.~2001, 2004; Petrosian 2001; Fujita et al.~2003;
Cassano \& Brunetti 2005).
Unavoidable $\gamma$-ray emission, due to the decay of the neutral 
pions generated through proton-proton collisions, is expected in 
the context of hadronic models (e.g. Blasi \& Colafrancesco 1999; Miniati 2003;
Pfrommer \& Ensslin 2004). 
Some $\gamma$-ray emission is also expected from those re-acceleration 
models that account for 
the general situation where both relativistic protons 
and electrons (including secondaries) interact with MHD turbulence
(Brunetti \& Blasi 2005; Brunetti 2008). Those halos with 
very steep spectrum are suitable targets to
constrain models and favour a turbulent re-acceleration 
scenario (\eg Brunetti et al.~2008): in fact it must be admitted that 
clusters hosting radio halos with very-steep spectrum should contain a very 
large population of CRp assuming the hadronic scenario; 
this also implies an unavoidably
large $\gamma$-ray emission from these clusters.

Only upper limits to the $\gamma$-ray emission from clusters have
been obtained so far (Reimer et al.~2003; Aharonian et al.~2009b), 
implying in some cases a fairly stringent 
constraint to the energy density of CRp, $<$10 \% of that of the thermal ICM
(Aharonian et al.~2009a). The FERMI/GLAST telescope will shortly provide 
more stringent constraints to the $\gamma$-ray 
properties of clusters and to the
energy density of CRp. 

\noindent
The radio halo in A2256 is the nearest steep-spectrum halo, and we show 
that the incoming FERMI/GLAST data will provide a prompt test of the 
hadronic scenario
and allow for constraining the cluster-magnetic field.

\noindent
A $\Lambda$CDM cosmology 
($H_{o}=70\,\rm km\,\rm s^{-1}\,\rm Mpc^{-1}$, $\Omega_{m}=0.3$,
$\Omega_{\Lambda}=0.7$) is adopted.

\section{The cluster Abell 2256}

Abell 2256 is a massive galaxy cluster at z=0.058, with 
0.1--2.4 keV X-ray luminosity $L_X \simeq 3.8 \cdot 10^{44}$erg/s (\eg
Ebeling et al.~1996).
The dynamical state of A2256 is complex and is thought to 
consist of at least three merging systems based on optical velocity 
dispersion (Berrington et al. 2002; Miller et al. 2003). 
A complex dynamics is also suggested by X--ray observations 
that revealed two separate peaks in the X-ray 
surface brightness distribution corresponding to the primary cluster 
and to the secondary subcluster, that is infalling onto the primary 
from the north--east (Briel et al. 1991; Sun et al. 2002).

\noindent
Radio observations revealed complex diffuse emission on large scale 
(Bridle et al. 1979; Rottgering et al. 1994; 
Clarke \& Ensslin 2006; Brentjens 2008) that consists of a 
bright relic, north--west of the cluster center, and of a fainter 
steep-spectrum Mpc--scale radio halo in the cluster central region.
Deep observations at 1400 and 300 MHz detect diffuse radio-halo 
emission up to a distance from cluster center 
$\sim 1.5 \, r_c \approx 520$ kpc (Clarke \& Ensslin 2006;
Brentjens 2008). A detailed spectral analysis of the halo emission 
derived a integrated spectral index between 0.3--1.4 GHz, $\alpha = 1.61$ 
($F(\nu) \propto \nu^{-\alpha}$), once the 
contribution from the embedded discrete radio sources is 
subtracted (Brentjens 2008).

\begin{figure}
\begin{center}
\includegraphics[width=7cm]{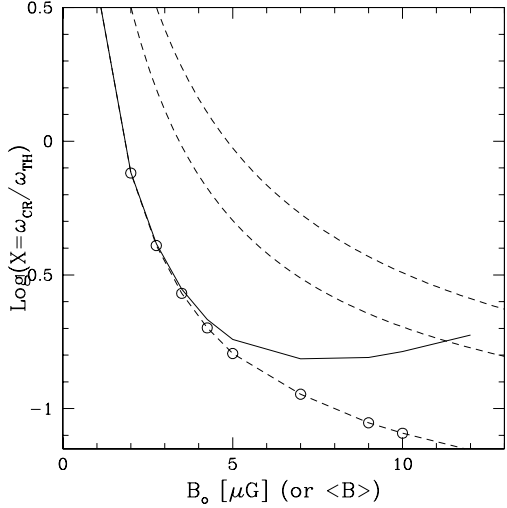}
\caption[]{
Ratio between relativistic CRp and thermal energy 
densities (for $r \leq 1.5 r_c$) for the {\it Steep} (dashed line, 
open circles) and {\it Flat} models (red dashed lines) as 
a function of $<B>$ ({\it Steep} model) 
and $B_o$ ({\it Flat} model; b=0.5,1 from bottom to top). 
The ratio between CRp$+B$ and thermal energy densities 
is shown for the {\it Steep} model (solid line).}
\label{Energy_CR}
\end{center}
\end{figure}

\begin{figure*}
\begin{center}
\includegraphics[width=7cm]{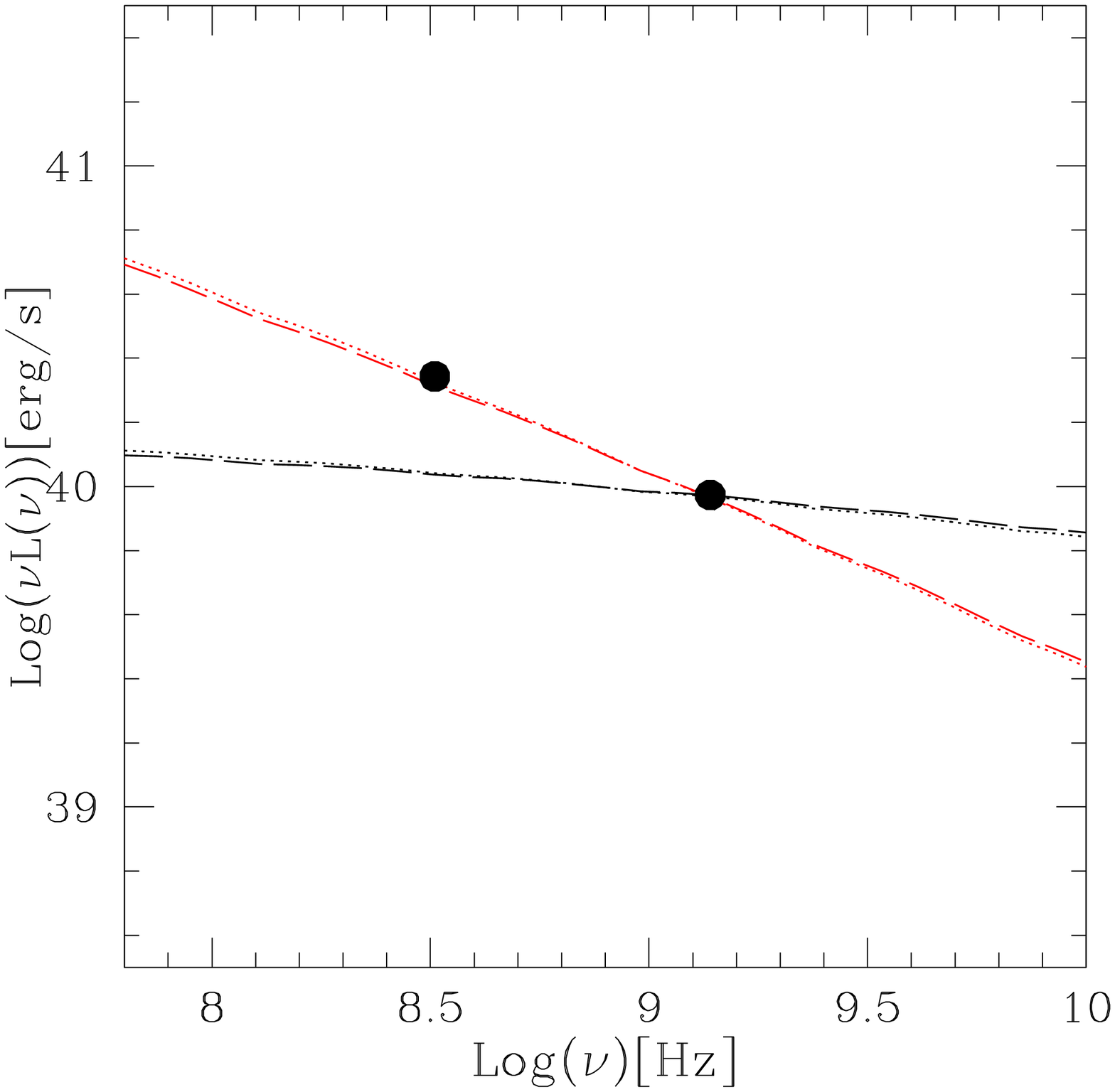}
\includegraphics[width=7cm]{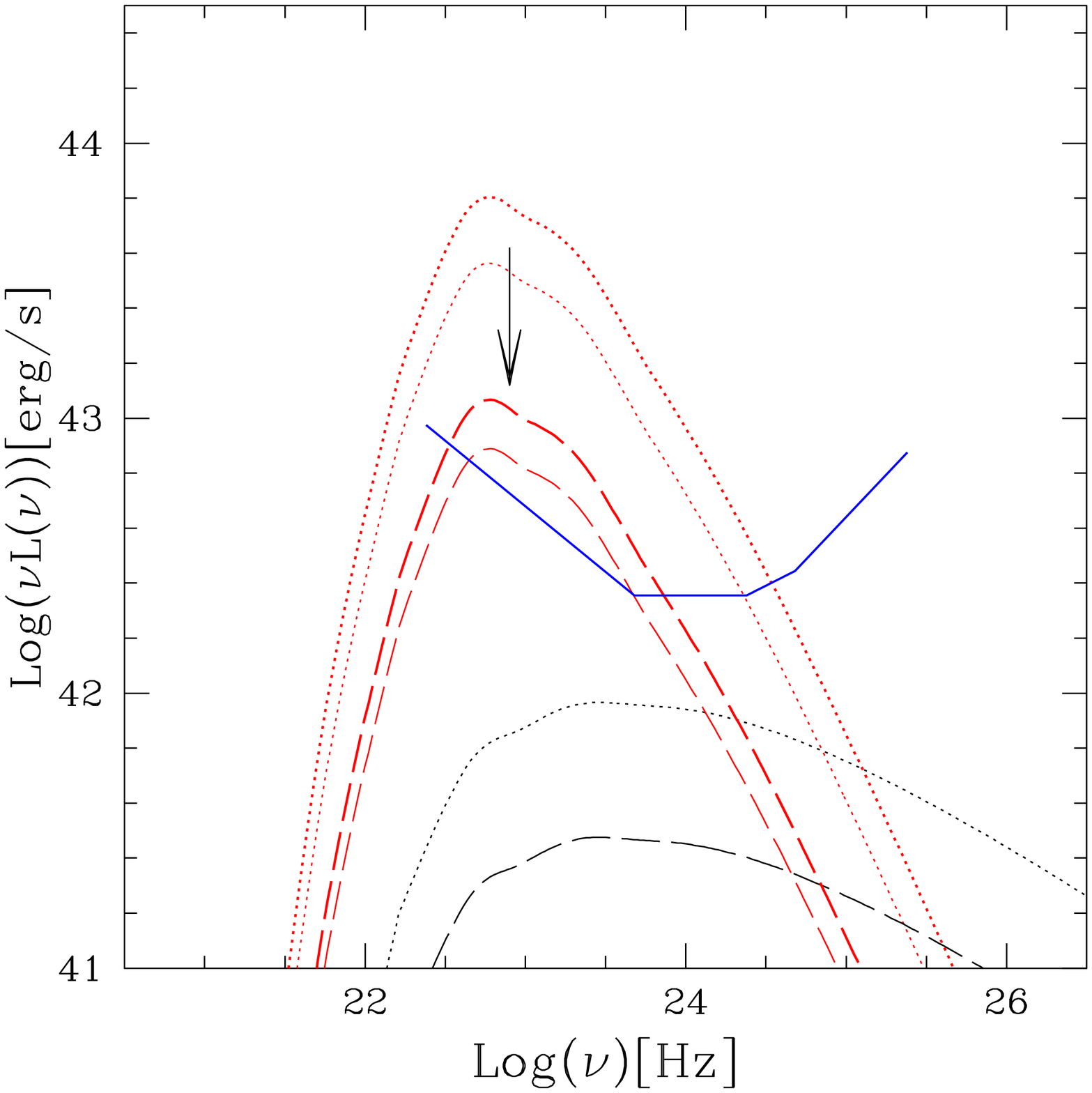}
\caption[]{Radio (left) and ($\pi^o$--decay) $\gamma$--rays from 
$r \leq 3 r_c$ (right) from A2256 assuming $b$=0.5 and $B_o$=2.65 (dotted lines) and
10 $\mu$G (dashed lines). Calculations are shown for $s=3.5$ (red lines)
in the case of the {\it Flat} (thick lines) and {\it Steep} model (thin lines).
Results for a {\it Steep} model with $s=2.4$ are also shown for comparison.
Monochromatic radio luminosities at 330 and 1400 MHz (filled points), the
EGRET upper limit (arrow) and the FERMI/GLAST reference sensitivity (solid--blue line) are
also shown.}
\label{Radio-Gamma}
\end{center}
\end{figure*}

\section{Hadronic models: formalism}

The decay chain that we consider for the injection
of secondary particles in the ICM due to p-p
collisions is (Blasi \& Colafrancesco 1999):

$$p+p \to \pi^0 + \pi^+ + \pi^- + \rm{anything}$$
$$\pi^0 \to \gamma \gamma$$
$$\pi^\pm \to \mu + \nu_\mu ~~~ \mu^\pm\to e^\pm \nu_\mu \nu_e.$$

\noindent
that is a threshold reaction that requires protons with kinetic
energy larger than $T_p \approx 300$ MeV. 

\noindent
The injection rate of pions is :

\begin{equation}
Q_{\pi}^{\pm,o}(E_{\pi^{\pm,o}},t)= n_{th} c 
\int_{p_{*}} dp N_p(p,t) \beta_p {{ F_{\pi}(E_{\pi},E_p) 
\sigma^{\pm,o}(p)}\over
{\sqrt{1 + (m_pc/p)^2} }},
\label{q_pi}
\end{equation}

\noindent
where $n_{th}$ is the number density of thermal protons,
and $F_{\pi}$ is the spectrum of pions from the collision between a CRp 
of energy $E_p$ and thermal protons (taken from Brunetti \& Blasi
2005). 
The inclusive cross section, 
$\sigma(p)$, is taken from the fitting formulae in Dermer (1986b) which 
allow to describe separately the rates of generation of 
$\pi^{-}$, $\pi^{+}$ and 
$\pi^{o}$, and $p_* = \max \{ p_{_{tr} },\, p_{\pi} \}$; $p_{tr}$ is the
threshold momentum of protons.

\noindent
The spectrum of $\gamma$-rays produced by the decay of the secondary 
$\pi^o$ is (\eg Dermer 1986ab; Blasi \& Colafrancesco 1999):

\begin{equation}
Q_{\gamma}(E_{\gamma})
= 2 \int_{E_{min}}^{E_p^{max}}
{{ Q_{\pi^o}(E_{\pi^o}) }\over{
\sqrt{E_{\pi}^2 - m_{\pi}^2 c^4} }}
d E_{\pi}
\label{gamma}
\end{equation}

\noindent
where $E_{min} = E_{\gamma} + {1/4} m_{\pi}^2 c^4 / E_{\gamma}$.

Charged pions decay into muons and secondary pairs (electrons and positrons).
Under the assumption that secondaries are not accelerated by other 
mechanisms, their spectrum approaches a stationary distribution due to
the competition between injection and energy losses 
(\eg Dolag \& Ensslin 2000) :

\begin{equation}
N_e^{\pm}(p)=
{1 \over
{\Big| \left( {{dp}\over{dt}} \right)_{\rm loss} \Big| }}
\int_{p}^{p_{\rm max}}
Q_e^{\pm}(p) dp.
\label{sec_stat}
\end{equation}

\noindent
where $Q_e^{\pm}$ is the injection rate of secondaries (e.g. Blasi \&
Colafrancesco 1999; Moskalenko \& Strong 1998), and 
radiative losses, that dominate for $\gamma > 10^3$ electrons in the ICM, 
are (\eg Sarazin 1999):

\begin{equation}
\Big| \left( {{ d p }\over{d t}}\right)_{\rm loss} \Big|
\simeq 3.3 \times 10^{-32} %\Big\{
\big( {{p/m_e c}\over{300}} \big)^2 \left[ \left( {{ B_{\mu G} }\over{
3.2}} \right)^2 + (1+z)^4 \right] 
\label{loss}
\end{equation}

\noindent
Assuming a power law distribution of CRp, $N_p(p) = K_p p^{-s}$, 
the spectrum of secondaries at high energies, 
$\gamma > 10^3$, is $N_e(p) \propto p^{-(s+1)} {\cal F}(p)$; 
${\cal F}$ accounts for the Log--scaling 
of the p-p cross section at high energies and makes the spectral shape 
slightly flatter than $p^{-(s+1)}$ (\eg Brunetti \& Blasi 2005).
The synchrotron spectrum from secondary e$^{\pm}$ is 
(e.g. Ribicky \& Lightman 1979):

\begin{eqnarray}
J_{syn}(\nu) = \sqrt{3} {{e^3}\over{m_e c^2}} B 
\int_0^{\pi/2} d\theta sin^2\theta \int dp N_e(p) 
F\big( {{\nu}\over{\nu_c}} \big)
\nonumber\\
\simeq C(\alpha, T) X n_{th}^2 
{{B^{1+\alpha} }\over{B^2 + B_{cmb}^2}} \nu^{-\alpha}
\label{jsyn}
\end{eqnarray}

\noindent where $C$ is a constant, $X=\omega_{CR}/\omega_{TH}$ is the
ratio between the energy densities of CRp and thermal protons, 
$F$ is the synchrotron
Kernel, $\nu_c$ is the critical frequency;
$\alpha \simeq s/2 -\Delta$, $\Delta \sim 0.1-0.15$ due to the Log-scaling 
of the cross section.

\section{Results}

In this Section we show that the steep spectrum of the halo in 
A2256 (Sect.2) allows for a prompt test of hadronic models and to 
constrain the magnetic field in the ICM.

\noindent 
We assume that the radio halo is due to synchrotron emission 
from scondary electrons, 
in which case the observed synchrotron spectral index, $\alpha =1.61$, implies
$s=3.4-3.5$. 
Parameters for the thermal ICM distribution in A2256, 
$n_o$, $T$, $r_c$, and $\beta$, are derived from Henry 
et al.~(1993) and Myers et al.~(1997). 

\noindent
We first adopt a {\it Steep} model that assumes a constant ratio between 
the CRp and energy density of thermal protons, 
$\omega_{CR}/\omega_{TH} = X$, and model the halo region with a 
homogeneus sphere with radius $R_H \sim 1.5 r_c$ and a volume 
averaged field (weighted for the synchrotron
emissivity) $<B>$.
The value of $X$, considering relativistic CRp only, 
that is requested to match the observed synchrotron spectrum
is shown in Figure 1 as a function of $<B>$.
We find that $<B> \leq 2 \mu$G can be excluded since the CRp
energy density becomes larger than the cluster thermal budget.
For stronger magnetic fields, $<B> \, \geq 4-5 \mu$G, 
$\omega_{CR} \leq 0.2 \omega_{TH}$ and the non-thermal
component becomes magnetically dominated.
The non-thermal energy content reaches a minimum, 
$\sim$0.16$\omega_{TH}$, for $<B> \approx 7-9 \mu$G that marks 
the minimum energy condition for hadronic models (Pfrommer \& Ensslin 2004). 
If we do not restrict to relativistic CRp 
and include also sub-relativistic CRp, due to the very steep spectrum,
the required energy budget is much larger than that in Figure 1, 
$\omega_{CR} \propto p_{min}^{-s+3}$, making the energetics
of CRp considerably larger.

\noindent We assume a spatial profile of 
the magnetic field $B= B_o \big( {{n_{th}}\over{n_o}} \big)^b$ 
(\eg Govoni \& Feretti 2004)
and find that the {\it Steep} model produces a radio-halo brightness
profile that drops by a factor 25--40 at $r \sim 1.5 \, r_c$, by 
adopting $b$=0.5--1 and $B_o \geq 5 \mu$G.
This is not consistent with the observed profile that drops, at the same
distance, by only a factor 5--8 (Clarke \& Ensslin 2006; Brentjens 2008).
Thus we consider a {\it Flat} hadronic model, with $\omega_{CR}$=const up to
$r \sim 1.5 \, r_c$ and $X$=const for larger $r$, 
that produces a drop of the brightness 
of a factor 8--12 at $r \sim 1.5 \, r_c$ for the range
of ($b$, $B_o$) given above; this is our {\it reference} model.
The energy request of the {\it Flat} hadronic model is also reported 
in Figure 1 considering the conservative case of relativistic
CRp only.
The large energy budget for the non-thermal components 
is a drawback of an hadronic origin of the radio halo in A2256.

This large budget and the steep spectrum of CRp imply an unavoidably 
efficient production of $\gamma$--rays at 0.1--1 GeV due to $\pi^o$ decay.
Consequently FERMI/GLAST observations provide an 
efficient and complementary way to test a hadronic origin of the halo.

\noindent In Figure 2 we show the expected radio (left) and 
$\gamma$-ray (right) spectra of A2256 
for different values of $B_o$ (see caption) (models anchored to the observed
1.4 GHz emission); we also report the case of 
a hadronic model with $s=2.4$.

\noindent We find that assuming a hadronic origin of the radio halo and 
adopting the appropriate spectrum of CRp,
the $\gamma$--ray upper limit obtained with EGRET observations 
(Reimer et al. 2003) already constrains $B_o > 2.5 \mu$G. 
Most important FERMI/GLAST should be able to detect Abell 2256 in the 
next years, provided that $B_o \leq 10-15 \mu$G.
This is highlighted in Figure 3 where we show the expected photon number 
with $E_{\gamma} \geq 100$ MeV as a function of $B_o$ in the case of both
{\it Steep} and {\it Flat} hadronic models.

\section{Discussion and Conclusions}

Radio halos have typical synchrotron spectral indices $\alpha \sim 1.2-1.3$
(\eg Ferrari et al. 2008), yet halos with steeper spectrum 
might be more common in the Universe (\eg Cassano et al. 2006) and 
present observations at GHz frequencies may select preferentially those halos with flatter spectrum. The discovery of a few radio halos with spectral index 
$\alpha > 1.5-1.6$ suggests that the emitting electrons are accelerated by rather inefficient mechanisms, \eg turbulent acceleration, and constrains models, such as the hadronic one, that would require very large energy budget to explain these sources (\eg Brunetti et al.2008).

\noindent A2256 hosts the nearest radio halo with steep spectrum, $\alpha = 1.61$, that would request a spectral slope of CRp $s=3.4-3.5$ adopting the hadronic scenario; in this case only a small fraction of the total energy-budget of supra-thermal CRp is expected to be associated with relativistic CRp.
We exploit two approaches based on hadronic models : a {\it Steep}
model that assumes a constant fraction $X=\omega_{CR}/\omega_{TH}$ 
and a {\it Flat} model that assumes $\omega_{CR}$=const in the halo
volume and $X$=const outside. The last one is our {\it reference} model
since the observed halo--brightness profile of A2256 in fact implies a rather
flat spatial distribution of CRp.

\noindent  
Even by considering only relativistic CRp, the hadronic model requires 
a large CRp-energy budget to explain the halo 
in A2256 for central fields $B_o < 10 \mu$G. This is a drawback 
of the hadronic scenario and also implies that 
the expected $\gamma$--ray luminosity of A2256 would be 10-20 times 
that of similar clusters hosting halos with the same radio 
luminosity but with $\alpha \sim 1.2$. 
Under these conditions we show that FERMI/GLAST should be able to detect A2256.

\noindent
Non detection would imply either that the halo is not of hadronic origin, 
or that the magnetic field in the central cluster region is 
$B_o \geq 10-15 \mu$G. In the latter case however we would 
admit the ad hoc possibility that A2256 is a cluster with unusually 
strong magnetic field since strong fields are presently 
observed only in cool-core clusters (\eg Carilli \& Taylor 2002; Govoni \&
Feretti 2004); future observations of Faraday Rotation will provide 
complementary information on the cluster-magnetic field.

\noindent On the other hand, detection of steep--spectrum $\gamma$--ray emission 
from A2256 would imply a hadronic origin of the halo, allowing also for 
an unprecedented measure of the magnetic field strength in the cluster. This will also suggest that very unusual acceleration mechanisms operate in the ICM, channeling a large fraction of the cluster energy into CRp with very steep spectrum.

\noindent If the halo is generated by turbulent re-acceleration the maximum $\gamma$-ray luminosity that is expected from A2256 can be estimated by requiring that the emission from secondaries matches the radio flux at the highest frequencies and is much smaller than that at lower frequencies (assumed to be dominated by re-accelerated electrons) (Reimer et al.2004; Donnert et al. 2009). This gives a $\gamma$-ray luminosity similar to that of the model with $s=2.4$ in Figure 2 implying that detection would be possible only for weak fields, $B_o < 1 \mu$G, with the $\gamma$-ray spectrum much flatter than in the case of a hadronic origin of the halo.

\begin{figure}
\begin{center}
\includegraphics[width=7cm]{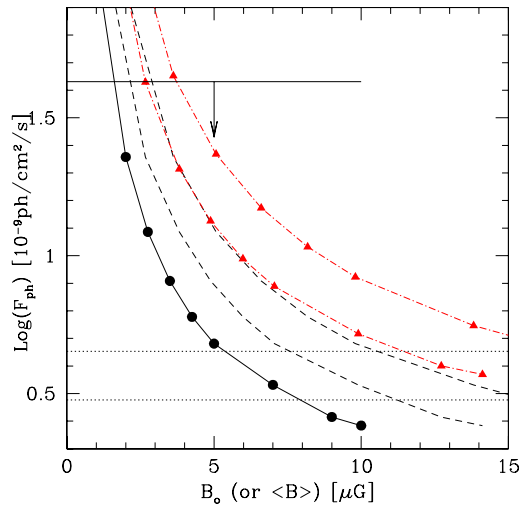}
\caption[]{Photon fluxes for $>$ 100 MeV are shown as a function of $B_o$ assuming {\it Flat} (red dot-dashed lines) and {\it Steep} models (dashed lines)
with $b$=0.5 and 1 (bottom to top). Photon flux vs $<B>$ for the {\it Steep} model is also shown (solid line with points). EGRET upper limit (arrow) and FERMI/GLAST sensitivity-range (dotted lines) are shown.}
\label{Lgamma_ph}
\end{center}
\end{figure}

\begin{acknowledgements}
This work is partially supported by grants PRIN-INAF2007 and 2008, 
and ASI-INAF I/088/06/0.
\end{acknowledgements}


\begin{thebibliography}{} 
\bibitem{} Aharonian F.A., et al., 2009a, A\&A 495, 27
\bibitem{} Aharonian F.A., et al., 2009b, A\&A 502, 437
%\bibitem{} Badhwar G.D, Golden R.L., Stephens S.A., 1977, Phy.Rev.D 15, 820
\bibitem{} Berezinsky V.S., Blasi P., Ptuskin V.S., 1997, ApJ 487, 529
\bibitem{} Berrington R.C., Lugger P.M., Cohn H.N., 2002, AJ 123, 2261
\bibitem{} Blasi P., 2001, APh 15, 223
\bibitem{} Blasi P., Colafrancesco S., 1999, APh 12, 169
\bibitem{} Blasi P., Gabici S., Brunetti G., 2007, Int. J. Mod. Phys. A 22, 
681
\bibitem{} Brentjens M.A., 2008, A\&A 489, 69
\bibitem{} Bridle, A.H., Fomalont E.B., Miley G.K., Valentijn E.A., 1979,
A\&A 80, 201
\bibitem{} Briel U.G., et al., A\&A 246, L10
\bibitem{} Brunetti G., 2008, ArXiv 0810.0692
\bibitem{} Brunetti G., Setti G., Feretti L., Giovannini G., 2001, MNRAS 320, 365
\bibitem{} Brunetti G., Blasi P., Cassano R., Gabici S., 2004, MNRAS 350, 1174
\bibitem{} Brunetti G., Blasi P., 2005, MNRAS 363, 1173
\bibitem{} Brunetti G., Lazarian A., MNRAS 378, 245
\bibitem{} Brunetti G., et al. 2008, Nature 455, 944
\bibitem{} Carilli C.L., Taylor G.B., 2002, ARA\&A 40, 319
\bibitem{} Cassano R., Brunetti G. 2005, MNRAS 357, 1313
\bibitem{} Cassano R., Brunetti G., Setti G., 2006, MNRAS 369,1577
\bibitem{} Clarke T.E., Ensslin T.A., 2006, ApJ 131, 2900
\bibitem{} Dennison B., 1980, ApJ 239, L93
\bibitem{} Dermer C.D., 1986a, A\&A 157, 223
\bibitem{} Dermer C.D., 1986b, ApJ 307, 47
\bibitem{} Dolag K., Ensslin T.A., 2000, A\&A 362, 151
\bibitem{} Dolag K., Grasso D., Springel V., Tkachev I., 2005, JCAP 1, 9
\bibitem{} Donnert J., Dolag K., Brunetti G., Cassano R., Bonafede A., 2009,
ArXiv:0905.2418
\bibitem{} Ebeling H., Voges W., B\"ohringer H., Edge A.C., Huchra J.P.,
Briel U.G., 1996, MNRAS 281, 799
\bibitem{} Ensslin T.A., Biermann P.L., Kronberg P.P., Wu X.-P., 1997,
ApJ 477, 560
\bibitem{} Ensslin T.A., Biermann P.L., Klein U., Kohle S., 1998, A\&A 332,
395
\bibitem{} Ferrari F., Govoni F., Schindler S. et al. 2008, SSRv 134, 93
\bibitem{} Fujita Y., Takizawa M., Sarazin C.L., 2003, ApJ 584, 190
\bibitem{} Gabici S., Blasi P., 2003, ApJ 583, 695
\bibitem{} Govoni F., Feretti L., 2004, Int. J. Mod. Phys. D 13, 1549
\bibitem{} Henry J.P., Briel U.G., Nulsen P.E.J., 1993, A\&A 271, 413
\bibitem{} Miller N.A., Owen F.N., Hill J.M., 2003, AJ 125, 2393
\bibitem{} Miniati F., 2003, MNRAS 342, 1009
\bibitem{} Moskalenko I.V., Strong A.W., 1998, ApJ 493, 694
\bibitem{} Myers S.T., Baker J.E., Readhead A.C.S., Leitch E.M., 1997, ApJ 485, 1
\bibitem{} Petrosian V., 2001, ApJ 557, 560 
\bibitem{} Pfrommer C., En\ss lin T. A. 2004, A\&A 413, 17
\bibitem{} Pfrommer C., Springel V, En\ss lin T.A., Jubelgas M., 2006,
MNRAS 367, 113
\bibitem{} Reimer A., Reimer O., Schlickeiser R., Iyudin A., 2004, A\&A 424, 773
\bibitem{} Reimer O., Pohl M., Sreekumar P., Mattox J.R., 2003, ApJ 588, 155
\bibitem{} Ribicky G.B., Lightmann A.P., {\it Radiative Processes in Astrophysics}, New York, Wiley-Interscience, 1979
\bibitem{} R\"ottgering H, Snellen I., Miley G., de Jong J.P., Hanisch R.J.,
Perley R., 1994, ApJ 436, 654
\bibitem{} Ryu D., Kang H., Hallman E., Jones T.W., 2003, ApJ 593, 599
\bibitem{} Ryu D., Kang H., Cho J., Das S., 2008, Science 320, 909
%\bibitem{} Stecker F.W., 1970, Ap\&SS 6, 377
%\bibitem{} Stephens S.A., Badhwar G.D., 1981, Ap\&SS 76, 213
\bibitem{} Sun M., Murray S.S., Markevitch M., Vikhlinin A., 2002, ApJ 565,
867
\bibitem{} V\"olk H.J., Aharonian F.A., Breitschwerdt D., 1996, SSRv 75, 279
\end{thebibliography}
\end{document}